\documentclass[12pt,a4paper]{article}

\usepackage{latexsym,amssymb,amsmath,amsbsy,epsfig,bbm}
\usepackage[numbers,sort&compress]{natbib}

\newcommand{\be}{\begin{equation}}
\newcommand{\ee}{\end{equation}}
\newcommand{\ba}{\begin{eqnarray}}
\newcommand{\ea}{\end{eqnarray}}
\newcommand{\bs}{\begin{subequations}}
\newcommand{\es}{\end{subequations}}

\newcommand{\Z}{\mathbb{Z}}

\begin{document}
\renewcommand{\thefootnote}{\fnsymbol{footnote}}

\title{
\normalsize \hfill CFTP/14-002
\\
\normalsize \hfill UWThPh-2014-5
\\*[7mm]
\LARGE Residual $\Z_2 \times \Z_2$ symmetries and lepton mixing}

\author{
L.\ Lavoura$^{(1)}$\thanks{E-mail: {\tt balio@cftp.ist.utl.pt}} \
and P.O.\ Ludl$ \, ^{(2)}$\thanks{E-mail: {\tt patrick.ludl@univie.ac.at}}
\\[3mm]
$^{(1)} \! $
\small Universidade de Lisboa, Instituto Superior T\'ecnico, CFTP, \\
\small 1049-001 Lisboa, Portugal
\\[1mm]
$^{(2)} \! $
\small University of Vienna, Faculty of Physics,\\
\small Boltzmanngasse 5, A--1090 Vienna, Austria
\\[3mm]
}

\date{20 January 2014}

\maketitle

\begin{abstract}
We consider two novel scenarios
of residual symmetries of the lepton mass matrices.
Firstly we assume a $\Z_2 \times \Z_2$ symmetry $G_\ell$
for the charged-lepton mass matrix
and a $\Z_2$ symmetry $G_\nu$ for the light neutrino mass matrix.
With this setting,
the moduli of the elements of one column of the lepton mixing matrix
are fixed up to a reordering.
One may interchange the roles of $G_\ell$ and $G_\nu$ in this scenario,
thereby constraining a row,
instead of a column,
of the mixing matrix.
Secondly we assume a residual symmetry group $G_\ell \cong \Z_m$ ($m>2$)
which is generated by a matrix with a doubly-degenerate eigenvalue.
Then,
with $G_\nu \cong \Z_2\times\Z_2$
the moduli of the elements of a row of the lepton mixing matrix get fixed.
Using the library of small groups
we have performed a search for groups
which may embed  $G_\ell$ and $G_\nu$ in each of these two scenarios.
We have found only two phenomenologically viable possibilities,
one of them constraining a column and the other one a row of the mixing matrix.
\end{abstract}

\newpage

\renewcommand{\thefootnote}{\arabic{footnote}}

\section{Introduction}

A group-theoretical philosophy
for explaining the phenomenological values of the lepton mixing parameters
has emerged during the last few years~\cite{lam0,lam,lam1,ge,yin,hagedorn0,
hagedorn2,he,smirnov,feruglio,lindner,hu,Walter,review-King,joshipura,hagedorn1}.
In that philosophy,
those values follow from the distinct Abelian symmetry groups---$G_\ell$
and $G_\nu$---under which the lepton mass matrices---$M_\ell$ and $M_\nu$,
respectively---are invariant.\footnote{The possibilities
for the experimental investigation of the implications of residual symmetries
are discussed in refs.~\cite{Luhn,Meloni,Ballett,Hanlon}.
Furthermore,
residual symmetries have also been considered in the quark sector~\cite{Araki}.}
Those matrices are defined by the mass terms
\be
\mathcal{L}_\text{mass} = - \bar \ell_L M_\ell \ell_R
+ \frac{1}{2}\, \nu_L^T M_\nu C^{-1} \nu_L + \mathrm{H.c.},
\ee
where $\ell_{L,R}$ are the left- and right-handed charged-lepton fields,
$\nu_L$ are the light neutrino fields,
and $C$ is the charge-conjugation matrix in Dirac space.
(We assume the neutrinos to be Majorana particles.)
Let $H_\ell \equiv M_\ell M_\ell^\dagger$;
if the mass matrices are diagonalized as
$U_\ell^\dagger H_\ell U_\ell
= D_\ell \equiv \mathrm{diag} \left( m_e^2,\, m_\mu^2,\, m_\tau^2 \right)$
and $U_\nu^T M_\nu U_\nu = D_\nu
\equiv \mathrm{diag} \left( m_1,\, m_2,\, m_3 \right)$,
then the lepton mixing matrix is given by $U_\text{PMNS} = U_\ell^\dagger U_\nu$.
($m_{1,2,3}$ denote the three neutrino masses.)
Let the symmetry group $G_\ell$ be generated by a matrix $L$
such that $L^{-1} H_\ell L = H_\ell$.\footnote{In our search
in section~\ref{Gell_Kleingroup},
$G_\ell$ is generated by \emph{two}\/ matrices $L_1$ and $L_2$
instead of just one.}
If we choose a basis in which $L$ is diagonal
and if we assume that the diagonal matrix elements of $L$ are all distinct,
then this invariance forces $H_\ell$ to be diagonal.
Thus,
in that basis $U_\ell = \mathbbm{1}_3$
(up to a permutation of the charged leptons)
and $U_\text{PMNS} = U_\nu$.
($\mathbbm{1}_3$ denotes the $3 \times 3$ unit matrix.)
In the same basis,
let a generator $N$ of $G_\nu$
be a unitary $3 \times 3$ matrix
of order two and with two different eigenvalues,
\textit{i.e.}\ $N^2 = \mathbbm{1}_3$ but $N \neq \pm \mathbbm{1}_3$.
Such a matrix can always be written as
\begin{equation}
\label{C}
N = \gamma \left( \mathbbm{1}_3 - 2 u u^\dagger \right),
\end{equation}
where $\gamma = \pm 1$
and $u = \left( u_1,\, u_2,\, u_3 \right)^T$ is a normalized column vector,
\textit{viz.}\
$u^\dagger u
= \left| u_1 \right|^2 + \left| u_2 \right|^2 + \left| u_3 \right|^2 = 1$.
Invariance of $M_\nu$ under $N$ means that $N^T M_\nu N = M_\nu$.
Then,
it follows from $N u = - \gamma u$ that
$N^\ast ( M_\nu u ) = N^T ( M_\nu u )
= \left( N^T M_\nu N \right) ( N u ) = - \gamma \left( M_\nu u \right)$.
But,
the eigenvalue $-\gamma$
of $N^\ast$ is non-degenerate;
therefore,
$M_\nu u \propto u^\ast$.
Since $M_\nu U_\nu = U_\nu^\ast D_\nu$ and the neutrino masses are non-degenerate,
$u$ must be one of the columns of $U_\nu = U_\text{PMNS}$.
It thence follows that $|u_{1,2,3}|$ are,
up to a reordering of the charged leptons,
the moduli of the matrix elements of a column
(one may still choose which column)
of $U_\text{PMNS}$.

The above-mentioned philosophy assumes that
there is a \emph{finite}\/ discrete group $G$
which has both $G_\ell$ and $G_\nu$ as subgroups.\footnote{If
$G$ is not assumed to be finite (and small),
then $G_\ell$ and $G_\nu$ will be largely arbitrary
and the philosophy will have little predictive power.}
It tries to find a suitable $G$ such that
the ensuing $|u_{1,2,3}|$ agree with the phenomenological values
of the moduli of the matrix elements of one of the columns of $U_\text{PMNS}$.
This has been done in ref.~\cite{lam}
under the assumption that $G$ is a subgroup of $SU(3)$
of order smaller than 512.
In ref.~\cite{lindner} a more complete search has been undertaken,
wherein $G$ was assumed to be a subgroup of $U(3)$ of order less than 1536.
Both refs.~\cite{lam} and~\cite{lindner} assume $G$ to possess a faithful
three-dimensional irreducible representation.
In ref.~\cite{lindner}
it was moreover assumed that $G$ fully determines $U_\text{PMNS}$,
because its subgroup $G_\nu$ is generated by \emph{two}\/ commuting matrices
$N$ and $N^\prime$,
both of the form in eq.~(\ref{C})
but with two mutually orthogonal vectors $u$ and $u'$,
respectively.
(Thus,
$G_\nu \cong \Z_2 \times \Z_2$ instead of $G_\nu \cong \Z_2$.)
A variant of this philosophy has been employed
in refs.~\cite{hagedorn2,hagedorn1},
where the neutrino mass terms have been assumed to be of the Dirac type and,
correspondingly,
the matrix $N$ has been assumed to generate
a group $G_\nu \cong \Z_n$ with $n > 2$.

In this paper we report on two group searches
that we have undertaken
and which might hold promise of relevant results.
In the first search---in section~\ref{Gell_Kleingroup}---we have assumed that
$G_\ell \cong \Z_2 \times Z_2$
(instead of the usual choice $G_\ell \cong \Z_m$ with $m>2$)
and $G_\nu \cong \Z_2$.
In the second search---in section~\ref{secondsearch}---we have assumed that
$G_\nu \cong \Z_2 \times \Z_2$ and that $G_\ell \cong \Z_m$
but with a doubly-degenerate eigenvalue,
in such a way that a \emph{row}\/
(instead of a column)
of $U_\text{PMNS}$ gets fixed.
In section~\ref{rows}
the results of our searches are confronted
with the phenomenological values.
Section~\ref{conclusions} contains the conclusions of this work.

\section{Group searches}
\label{searches}

\subsection{First search: $G_\ell\cong \Z_2 \times \Z_2$, $G_\nu \cong \Z_2$}
\label{Gell_Kleingroup}

We consider in this section a scenario in which
the lepton flavour symmetry group $G$
is broken to two residual symmetry subgroups
$G_\ell \cong \Z_2 \times \Z_2$ and $G_\nu \cong \Z_2$.
The symmetry group $G_\ell$ holds in the charged-lepton sector
while $G_\nu$ holds in the neutrino sector.
We require the embedding group $G$ to be \emph{finite}\/
and to have \emph{a faithful three-dimensional irreducible representation
$D \left( G \right)$}.
We assume that $- \mathbbm{1}_3 \not\in D \left( G_\ell \right)$
and also $- \mathbbm{1}_3 \not\in D \left( G_\nu \right)$.
Furthermore,
there must be a mismatch between the residual symmetries $G_\ell$ and $G_\nu$,
\textit{i.e.} we require that $G_\nu \not\subset G_\ell$.

To summarize,
we have searched for groups $G$ which fulfil the following conditions:
\begin{enumerate}
\item $G$ is finite.
\item $G$ has a faithful three-dimensional irreducible representation
$D \left( G \right)$.
\item $G$ has two subgroups,
$G_\ell \cong \Z_2 \times \Z_2$ and $G_\nu \cong \Z_2$,
which have a trivial intersection,
\textit{i.e.}\ $G_\ell \cap G_\nu = \{e\}$.
\item Neither $D \left( G_\ell \right)$
nor $D \left( G_\nu \right)$ contain the matrix $- \mathbbm{1}_3$.
\end{enumerate}
Since we are interested in groups $G$
which have a $\Z_2 \times \Z_2$ subgroup,
$\mathrm{ord} \left( G \right)$ must be divisible by four.
Since we require $G$
to have a three-dimensional irreducible representation,
$\mathrm{ord} \left( G \right)$ must be divisible by three.
Thus,
we only need to consider groups of order divisible by 12.

Since $G$ is finite,
there is a basis in which $D \left( G \right)$ consists of unitary matrices.
Since $G_\ell \cong \Z_2 \times \Z_2$ is Abelian,
a basis can be chosen in which $D \left( G_\ell \right)$ 
is formed by diagonal matrices.
Thus,
$D \left( G_\ell \right)$ comprehends $\mathbbm{1}_3$ and
\bs
\label{AB}
\ba
L_1 &=& \alpha\ \mathrm{diag} \left( +1,\, -1,\, -1 \right),
\\
L_2 &=& \beta\ \mathrm{diag} \left( -1,\, +1,\, -1 \right),
\\
L_1 L_2 = L_2 L_1 \equiv L_3
&=& \alpha \beta\ \mathrm{diag} \left( -1,\, -1,\, +1 \right),
\ea
\es
where both $\alpha$ and $\beta$ may be either $+1$ or $-1$.
The residual symmetry $G_\ell$ means that
$L_1^{-1} H_\ell L_1 = L_2^{-1} H_\ell L_2 = H_\ell$.
Therefore,
in the basis where $L_1$ and $L_2$ are as in eqs.~(\ref{AB}),
$H_\ell$ must be diagonal.

In the same basis,
the generator $N$ of $D \left( G_\nu \right)$
is a unitary $3 \times 3$ matrix of order two,
\textit{i.e.}\ a matrix of the form in eq.~(\ref{C}),
where $u = \left( u_1,\, u_2,\, u_3 \right)^T$ is a normalized column vector.
Then,
up to a reordering,
the $|u_k|$ ($k=1,2,3$) are the moduli
of the matrix elements of one column of $U_\text{PMNS}$.
Given the matrices $L_1$,
$L_2$,
and $N$ \emph{in an arbitrary basis},
one may compute the $|u_k|^2$
without the need to diagonalize $L_1$ and $L_2$;
indeed,
\be
\label{uisquared}
\left| u_k \right|^2 = \frac{1}{4}
\left[ 1 + \frac{\mathrm{tr} \left( L_k N \right)}
{\mathrm{tr} \left( L_k \right) \mathrm{tr} \left( N \right)}
\right].
\ee
Equation~(\ref{uisquared}) is easily verified
in the basis where eqs.~(\ref{C}) and~(\ref{AB}) hold;
since it is written in terms of traces,
it holds in any other basis---even in one where $D \left( G \right)$
is not formed by unitary matrices.
One may thus compute the moduli of the matrix elements
of one column of $U_\text{PMNS}$ just from the knowledge of $L_1$,
$L_2$,
and $N$ in an arbitrary basis.\footnote{The computation
of mixing-matrix elements from invariant traces
was pioneered in ref.~\cite{branco}.}

The computer algebra system GAP~\cite{GAP}
has access to SmallGroups~\cite{SmallGroups},
a library of all the groups
(up to isomorphisms)
of order smaller than 2\,000---
excluding the 49\,487\,365\,422 groups of order 1\,024.
Since there are 408\,641\,062 groups of order $1536 = 12 \times 128$,
we have restricted our search to the 1\,336\,749 groups
of order $12 n$ for $n \leq 127$.
We have furthermore excluded groups $G$ which are direct products of the form
\begin{equation}
\label{directcyclic}
G \cong \Z_m \times G' \quad (m \ge 2),
\end{equation}
because such groups do not provide any restrictions
beyond those already following from the smaller group $G'$.

Going through these groups,
by constructing their character tables,
we have sieved out the groups which have a faithful
three-dimensional irreducible representation.
We have used the GAP package SONATA~\cite{sonata}
to find all the subgroups of the groups under investigation.
For those groups which have a $\Z_2 \times \Z_2$ subgroup
and a $\Z_2$ subgroup with trivial intersection,
we have explicitly constructed
all the non-equivalent faithful three-dimensional irreducible representations
$D$ and we have computed all the candidates
for pairs $\left( D \left( G_\ell \right),\, D \left( G_\nu \right) \right)$.
When neither $D \left( G_\ell \right)$ nor $D \left( G_\nu \right)$
contained $- \mathbbm{1}_3$,
we have computed the corresponding $\left| u_k \right|^2$
through eq.~(\ref{uisquared}).
The results can be found in table~\ref{results-groupsearch1}.

In table~\ref{results-groupsearch1}
(and in the second column of table~\ref{list})
one observes that,
whenever $G_\ell$ and $G_\nu$ together generate a group $D_n$ with even $n$,
this leads to $\left( \left| u_1 \right|^2,\, \left| u_2 \right|^2,\,
\left| u_3 \right|^2 \right) = \left( 0,\, \sin^2{\frac{2 \pi}{m}},\,
\cos^2{\frac{2 \pi}{m}} \right)$ with $m = 2n$ and,
possibly,
smaller (integer) values of $m$.\footnote{The group $D_{14}$
is of particular interest,
especially for quark mixing,
because it nicely fits Cabibbo mixing~\cite{hagedorn},
as can be seen in
the second line before the last
of table~\ref{results-groupsearch1}.}
The group $D_n$ may be defined as consisting of the matrices
\be
X(p) = \left( \begin{array}{cc}
- \cos{\left( p \alpha_n \right)} & - \sin{\left( p \alpha_n \right)} \\
- \sin{\left( p \alpha_n \right)} & \cos{\left( p \alpha_n \right)}
\end{array} \right)
\quad \mbox{and} \quad
Y(p) = \left( \begin{array}{cc}
\cos{\left( p \alpha_n \right)} & - \sin{\left( p \alpha_n \right)} \\
\sin{\left( p \alpha_n \right)} & \cos{\left( p \alpha_n \right)}
\end{array} \right),
\ee
where $\alpha_n \equiv 2 \pi / n$ and $p = 0, 1, 2, \ldots, n-1$.
For even $n$,
this group has a $\Z_2 \times \Z_2$ subgroup formed by $\mathbbm{1}_2$,
$Y(n/2)$,
$X(n/2)$,
and $X(0)$.
The group $D_n$ is a subgroup of $SO(3)$ through its \emph{reducible}\/
triplet representation
\be
X(p) \to \widetilde X(p) \equiv  \left( \begin{array}{cc}
- 1 & 0_{1 \times 2} \\
0_{2 \times 1} & X(p)
\end{array} \right),
\quad 
Y(p) \to \widetilde Y(p) \equiv \left( \begin{array}{cc}
1 & 0_{1 \times 2} \\
0_{2 \times 1} & Y(p)
\end{array} \right).
\ee
In this representation of $D_n$,
its $\Z_2 \times \Z_2$ subgroup is formed by
\be
\left\{
\mathbbm{1}_3,\,
\widetilde Y (n/2) = L_1,\,
\widetilde X (n/2) = L_2,\,
\widetilde X (0) = L_3
\right\},
\ee
where the matrices $L_{1,2,3}$ are as in eqs.~(\ref{AB})
with $\alpha = \beta = +1$.
The $G_\nu$ subgroup is formed by
\be
\left\{
\mathbbm{1}_3,\, \widetilde X(p)
\right\}.
\ee
By using eq.~(\ref{uisquared}) one then obtains $\left| u_1 \right|^2 = 0$
and $\left| u_2 \right|^2 = \sin^2{\left( p \alpha_n / 2 \right)}$.

\subsection{Second search: $G_\ell\cong \Z_n$, $G_\nu \cong \Z_2 \times \Z_2$}
\label{secondsearch}

One may interchange the roles
of Klein's four group $\Z_2 \times \Z_2$
and of the cyclic group $\Z_2$
in section~\ref{Gell_Kleingroup}.
When one does that,
the neutrino mass matrix $M_\nu$ is invariant under $\Z_2 \times \Z_2$,
\textit{viz.}\ $N_1^T M_\nu N_1 = N_2^T M_\nu N_2 = M_\nu$
with $N_1^2 = N_2^2 = \mathbbm{1}_3$.
If we choose the basis where $N_1$ and $N_2$ are diagonal,
then in that basis $M_\nu$ will be diagonal too.
Let $H_\ell$ possess a residual $\Z_2$ symmetry,
\textit{i.e.}\ $L^{-1} H_\ell L = H_\ell$
with $L = \gamma \left( \mathbbm{1}_3 - 2 u u^\dagger \right)$
as in equation~(\ref{C}).
Consequently,
$L H_\ell = H_\ell L$ and therefore $L ( H_\ell u )
= H_\ell L u = -\gamma \left( H_\ell u \right)$.
Then,
since the eigenvalue $-\gamma$ of $L$ is non-degenerate,
$H_\ell u \propto u$.
Now,
the eigenvalues of $H_\ell$,
\textit{viz.}\ the squares of the charged-lepton masses,
are non-degenerate.
Therefore,
$u$ must be a column of the unitary matrix $U_\ell$ diagonalizing $H_\ell$.
Since we are in the basis where $M_\nu$ is diagonal,
$U_\text{PMNS} = U_\ell^\dagger$ up to a permutation of the rows of $U_\text{PMNS}$.
We have thus found that in this case
the residual symmetries constrain \emph{a row},
rather than a column,
of the mixing matrix $U_\text{PMNS}$.
The possible restrictions on the moduli of the matrix elements of the row
are of course precisely the same
as those obtained in section~\ref{Gell_Kleingroup},
see table~\ref{results-groupsearch1}.

An important feature of the scenario just described is that
the matrix $L$ generating the residual symmetry group of $H_\ell$
has two degenerate eigenvalues and the third eigenvalue is different.
The matrix $L$ is,
however,
restricted by the condition $L^2 = \mathbbm{1}_3$,
since it generates a group $\Z_2$.
We now lift this restriction and
suppose instead that $L$ generates a group $\Z_n$ with $n > 2$,
\textit{i.e.}\ $L^n = \mathbbm{1}_3$.
We thus assume that in the neutrino sector
there is a residual symmetry $G_\nu \cong \Z_2 \times \Z_2$,
represented by $D \left( G_\nu \right)$ which,
in an appropriate basis,
is formed by $\mathbbm{1}_3$ together with
\bs
\label{ABprime}
\ba
N_1 &=& \alpha\ \mathrm{diag} \left( +1,\, -1,\, -1 \right),
\\
N_2 &=& \beta\ \mathrm{diag} \left( -1,\, +1,\, -1 \right),
\\
N_1 N_2 = N_2 N_1 \equiv N_3
&=& \alpha \beta\ \mathrm{diag} \left( -1,\, -1,\, +1 \right).
\ea
\es
In this basis,
$M_\nu$ is diagonal and therefore
$U_\text{PMNS} = U_\ell^\dagger$ up to a permutation of rows.
In the charged-lepton sector the residual symmetry is $\Z_n$,
generated by a matrix $L$ \emph{with a degenerate eigenvalue} $\sigma$
and another eigenvalue $\rho \neq \sigma$
(of course $\sigma^n = \rho^n = 1$).
Let,
in the basis where eqs.~(\ref{ABprime}) hold,
$v=\left( v_1,\, v_2,\, v_3 \right)^T$
denote the normalized eigenvector of $L$
corresponding to the eigenvalue $\rho$.
One may then write
\begin{equation}
\label{Cprime}
L = \sigma \mathbbm{1}_3 + \left( \rho - \sigma \right) v v^\dagger.
\end{equation}
The $\left| v_k \right|$ are,
up to a reordering,
the moduli of the matrix elements of one row of $U_\text{PMNS}$.
They may be computed in a basis-independent way through
\be
\label{visquared}
\left| v_k \right|^2 = \frac{1}{2 \left( \rho - \sigma \right)}
\left[ \rho - \frac{\mathrm{tr} \left( N_k L \right)}
{\mathrm{tr} \left( N_k \right) }
\right].
\ee

Thus,
we have searched for groups $G$ which fulfil the following conditions:
\begin{enumerate}
\item $G$ is finite.
\item $G$ has a faithful three-dimensional irreducible representation
$D \left( G \right)$.
\item $G$ has two subgroups,
$G_\ell \cong \Z_n$ ($n>2$) and $G_\nu \cong \Z_2 \times \Z_2$,
which have a trivial intersection,
\textit{i.e.}\ $G_\ell \cap G_\nu = \{e\}$.
\item $D \left( G_\nu \right)$ does not contain $-\mathbbm{1}_3$.
\item $D \left( G_\ell \right)$ is generated by a matrix $L$
which has a twice degenerate eigenvalue $\sigma$
and another eigenvalue $\rho$ which differs from $\sigma$.
\item The group $\left\langle\langle G_\ell,\, G_\nu \right\rangle\rangle$
generated by $D \left( G_\ell \right)$ and $D \left( G_\nu \right)$
is non-Abelian.
\end{enumerate}
Once again,
we have excluded groups of the form $G \cong \Z_m \times G'$ with $m\geq 2$.
For each group of order smaller than\footnote{We have stopped this search
at a rather low group order because
the construction of the irreducible representations
becomes, for large groups,
extremely expensive in terms of computer time.}
800 fulfilling the above requirements,
we have computed the corresponding $\left| v_k \right|^2$
by means of eq.~(\ref{visquared}).
The results can be found in table~\ref{results-groupsearchZn}.

\section{The case $\left( 1/4,\, 1/4,\, 1/2 \right)$}
\label{rows}

One sees in tables~\ref{results-groupsearch1} and~\ref{results-groupsearchZn}
that most predicted columns or rows of $U_\text{PMNS}$
contain a zero matrix element.
Such a situation is phenomenologically excluded\footnote{One might consider
the possibility where our predictions only hold as a first approximation
and are corrected by
other
effects---for instance,
suppressed terms in the Lagrangian
and/or the renormalization-group evolution of the parameters of $U_\text{PMNS}$.
We shall not entertain such possibilities here.}
and therefore most of the data in those tables
seem irrelevant for our purposes.

The remaining cases are more encouraging.
The possibility $\left( \left| u_1 \right|^2,\, \left| u_2 \right|^2,\,
\left| u_3 \right|^2 \right) = \left( \frac{3 + \sqrt{5}}{8},\, 
1/4,\, \frac{3 - \sqrt{5}}{8} \right)$,
in the last line of table~\ref{results-groupsearch1},
was recently discovered and constitutes a viable prediction
for the first column of $U_\text{PMNS}$~\cite{varzielas}.
On the other hand,
the possibility $\left( \left| u_1 \right|^2,\, \left| u_2 \right|^2,\,
\left| u_3 \right|^2 \right) = \left( 1/4,\, 1/4,\, 1/2 \right)$
gives a rather poor fit to the second column of $U_\text{PMNS}$.

Here we shall instead consider the case
$\left( \left| v_1 \right|^2,\, \left| v_2 \right|^2,\,
\left| v_3 \right|^2 \right) = \left( 1/4,\, 1/4,\, 1/2 \right)$
as a prediction for the third \emph{row}\/ of $U_\text{PMNS}$.
Using the standard parametrization for $U_\text{PMNS}$,
one then has
\bs
\ba
c_{23}^2 c_{13}^2 &=& 1/2,
\label{1} \\
s^2_{12} s^2_{23} + c^2_{12} c^2_{23} s^2_{13}
- 2 s_{12} c_{12} s_{23} c_{23} s_{13} \cos{\delta} &=& 1/4,
\label{2} \\
c^2_{12} s^2_{23} + s^2_{12} c^2_{23} s^2_{13}
+ 2 s_{12} c_{12} s_{23} c_{23} s_{13} \cos{\delta} &=& 1/4,
\label{3}
\ea
\es
where $s_i \equiv \sin{\theta_i}$ and $c_i \equiv \cos{\theta_i}$
for $i = 12, 13, 23$.

The sum of eqs.~(\ref{2}) and~(\ref{3}) is equivalent to eq.~(\ref{1}).
It makes a prediction for $\theta_{23}$ as a function of $\theta_{13}$:
\be
\label{23}
s_{23}^2 = \frac{1 - 2 s_{13}^2}{2 - 2 s_{13}^2}.
\ee
With $0.0169 \le s_{13}^2 \le 0.0315$
at $3\sigma$ level~\cite{fogli},
this yields $0.4837 \le s_{23}^2 \le 0.4914$.
This means that \emph{the atmospheric mixing angle
is maximal}\/ for all practical purposes.

\begin{figure}
\begin{center}
\epsfig{file=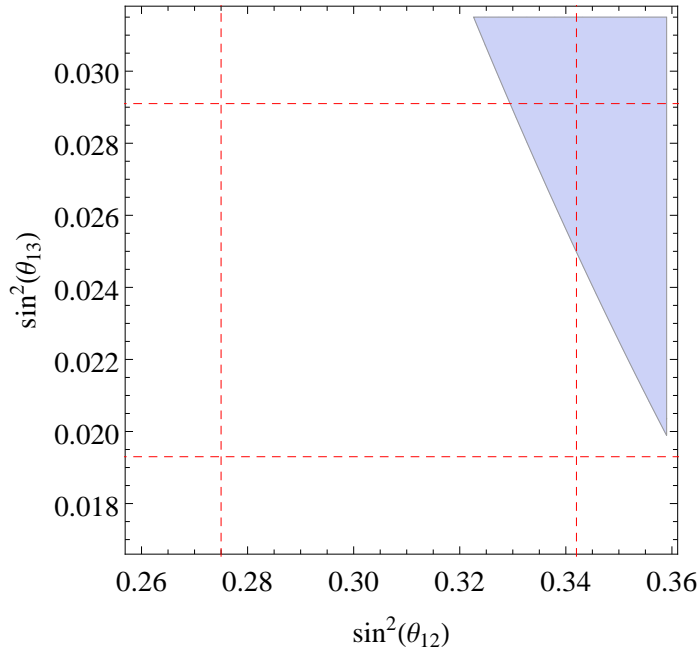,width=0.57\textwidth}
\end{center}
\caption{The area in the $\sin^2{\theta_{12}}$--$\sin^2{\theta_{13}}$ plane
allowed by our prediction in eq.~(\ref{bounds}).
The dotted lines represent
the $2\sigma$ phenomenological bounds of ref.~\cite{fogli}
on those parameters;
the shaded area extends to their $3\sigma$ bounds.}
\label{intervals}
\end{figure}

The difference between eqs.~(\ref{2}) and~(\ref{3})
yields a prediction for $\cos{\delta}$:
\be
4 s_{12} c_{12} s_{23} c_{23} s_{13} \cos{\delta} =
\left( s_{12}^2 - c_{12}^2 \right) \left( s_{23}^2 - c_{23}^2 s_{13}^2 \right).
\ee
Using eq.~(\ref{23}),
this gives
\be
\label{cos}
\cos{\delta} = - \frac{c_{12}^2 - s_{12}^2}{4 s_{12} c_{12}}\,
\frac{1 - 3 s_{13}^2}{\sqrt{s_{13}^2 - 2 s_{13}^4}}.
\ee
Since $c_{12} > s_{12}$,
$\cos{\delta}$ is predicted to be negative.\footnote{This is
not very meaningful because it just follows from our choice of fitting
$\left( \left| v_1 \right|^2,\, \left| v_2 \right|^2,\,
\left| v_3 \right|^2 \right) = \left( 1/4,\, 1/4,\, 1/2 \right)$
to the \emph{third}\/ row of $U_\text{PMNS}$.
If we had opted to fit it to the \emph{second}\/ row instead,
then the predicted value of $\cos{\delta}$ would be symmetric
to the one in eq.~(\ref{cos}).}
Moreover,
$\left| \cos{\delta} \right|$ is quite large;
the bound $\cos^2{\delta} \le 1$ gives
\be
\sin{\left( 2 \theta_{12} \right)} \ge \frac{1 - 3 s_{13}^2}{1 - s_{13}^2}
\approx 1 - 2 s_{13}^2 - 2 s_{13}^4 - 2 s_{13}^6 - \cdots.
\label{bounds}
\ee
This implies that $\theta_{12}$ and $\theta_{13}$
cannot be both  within their $1 \sigma$ intervals of ref.~\cite{fogli}
and can only marginally be both
within their $2 \sigma$ intervals, see fig.~\ref{intervals}.
Anyway,
the angle $\delta$ should be close to either $0$ or $\pi$,
\textit{i.e.}\ $CP$ violation in lepton mixing is predicted to be small.

\section{Conclusions}
\label{conclusions}

In this work,
using the software GAP and the SmallGroups Library,
we have looked for \emph{finite}\/ groups $G$ which have
{\em a faithful three-dimensional
irreducible representation $D \left( G \right)$}\/
and have two subgroups,
$\Z_n$ and $\Z_2 \times \Z_2$,
with a trivial intersection.
Moreover,
$D \left( \Z_n \right)$ should have {\em a twice degenerate eigenvalue}\/
and neither $D \left( \Z_n \right)$ (for $n=2$) nor $D \left( \Z_2 \times \Z_2 \right)$
should contain the matrix $- \mathbbm{1}_3$.
When $n=2$ we have taken the search up to group order 1536
but for $n>2$ we only reached group order 800.

Applying the results of our search to the prediction of lepton mixing,
we have noticed that almost all the groups that we have found
lead to a zero mixing matrix element,
which is phenomenologically disallowed.
There are only two exceptions.
In one of them,
the groups $[60, 5] \cong A_5$ and $[1080, 260] \supset [60, 5]$
may lead to the first column
of the lepton mixing matrix having elements with moduli squared
$\left( 0.6545,\, 0.25,\, 0.0955 \right)$;
this is viable and had already been found in a previous paper~\cite{varzielas}.
In the other exception,
many groups---see tables~1 and~2---may lead to either the second
or the third row of the lepton mixing matrix having elements with moduli
$\left( 1/2,\, 1/2,\, 1/\sqrt{2} \right)$;
the consequences of this prediction are
a very close to maximal atmospheric mixing angle
and $\left| \cos{\delta} \right|$ straddling 1.

\vspace*{5mm}

\paragraph{Acknowledgements:}
The work of LL is supported through
the Marie Curie Initial Training Network ``UNILHC'' PITN-GA-2009-237920
and also through the projects PEst-OE-FIS-UI0777-2013,
PTDC/FIS-NUC/0548-2012,
and CERN-FP-123580-2011
of the Portuguese {\it Funda\c c\~ao para a Ci\^encia e a Tecnologia}\/ (FCT).
POL acknowledges support through the Austrian Science Fund (FWF) via
project No.~P~24161-N16.

\begin{table}
\begin{small}
\begin{center}
\renewcommand{\arraystretch}{1.2}
\begin{tabular}{|l|l|l|}
\hline
$G$ &
$\left( \left| u_1 \right|^2,\, \left| u_2 \right|^2,\,
\left|u_3 \right|^2 \right)$ &
$\left\langle \left\langle G_\ell,\, G_\nu \right\rangle \right\rangle$
\\ \hline
$[24, 12];\ [96, 64];\ [168, 42];\ [216, 95];\ [384, 568];$ &
$\left( 0,\,\sin^2{\frac{2\pi}{8}},\,\cos^2{\frac{2\pi}{8}}\right)$ &
$[8, 3]$
\\
$[600, 179];\ [648, 259];\ [648, 260];\ [648, 266];$ &
$=\left( 0,\, 1/2,\, 1/2 \right)$ &
\\
$[648, 563];\ [864, 701];\ [1080, 260];\ [1176, 243] $ & &
\\ \hline
$[216, 95];\ [648, 259];\ [648, 260];\ [648, 266];$ &
$\left( 0,\,\sin^2{\frac{2\pi}{12}},\,\cos^2{\frac{2\pi}{12}}\right)$ &
$[12, 4]$
\\
$[648, 563];\ [864, 701]$ &
$=\left( 0,\, 1/4,\, 3/4 \right)$ &
\\ \hline
$[384, 568]$ &
$\left( 0,\,\sin^2{\frac{2\pi}{16}},\,\cos^2{\frac{2\pi}{16}}\right)$ &
$[16, 7]$ \\
 & $\approx \left( 0,\, 0.1464,\, 0.8536 \right)$ &
\\ \hline
$[600, 179]$ &
$\left( 0,\,\sin^2{\frac{2\pi}{10}},\,\cos^2{\frac{2\pi}{10}}\right)$ &
$[20, 4]$
\\
 & $\approx \left( 0,\, 0.3455,\, 0.6545 \right);$ &
\\
 & $\left( 0,\,\sin^2{\frac{2\pi}{20}},\,\cos^2{\frac{2\pi}{20}}\right)$ &
\\
 & $\approx \left( 0,\, 0.0955,\, 0.9045 \right)$ &
\\ \hline
$[864, 701]$ &
$\left( 0,\,\sin^2{\frac{2\pi}{24}},\,\cos^2{\frac{2\pi}{24}}\right)$ &
$[24, 6]$
\\
 & $\approx \left( 0,\, 0.0670,\, 0.9330 \right)$ &
\\ \hline
$[1176, 243]$ &
$\left( 0,\, \sin^2{\frac{2 \pi}{7}},\, \cos^2{\frac{2 \pi}{7}} \right)$ &
$[28, 3]$
\\
 & $\approx \left( 0,\,  0.6113,\,  0.3887 \right);$ &
\\
 & $\left( 0,\, \sin^2{\frac{2 \pi}{14}},\,
\cos^2{\frac{2 \pi}{14}} \right)$ &
\\
 & $\approx \left( 0,\, 0.1883,\, 0.8117 \right);$ &
\\
 & $\left( 0,\, \sin^2{\frac{2 \pi}{28}},\, \cos^2{\frac{2 \pi}{28}} \right)$ &
\\
 & $\approx \left( 0,\, 0.0495,\, 0.9505 \right)$ &
\\ \hline
$[24, 12];\ [96, 64];\ [168, 42];\ [216, 95];\ [384, 568];$ &
$\left( 1/4,\, 1/4,\, 1/2\right)$ &
$[24, 12]$
\\
$[600, 179];\ [648, 259];\ [648, 260];\ [648, 266];$ & &
\\
$[648, 563];\ [864, 701];\ [1080, 260];\ [1176, 243] $ & &
\\ \hline
$[60, 5];\ [1080, 260]$ &
$\left( 1/4,\, \frac{3-\sqrt{5}}{8},\, \frac{3+\sqrt{5}}{8} \right)$ &
$[60, 5]$
\\
 & $\approx \left( 0.25,\, 0.0955,\, 0.6545 \right)$ &
\\ \hline
\end{tabular}
\caption{In the first column,
the groups resulting from the search described in section~2.1;
the symbol $\left[ g,\, j \right]$
denotes the $j$-th group of order $g$ in the SmallGroups Library.
In the second column,
the corresponding values for the $\left| u_k \right|^2$
($k = 1, 2, 3$).
In the third column,
the symbol $\left\langle \left\langle G_\ell,\, G_\nu
\right\rangle \right\rangle$ denotes the group generated by
$D \left( G_\ell \right)$ and $D \left( G_\nu \right)$,
\textit{i.e.}\
the smallest finite group having $G_\ell$ and $G_\nu$
as subgroups.
A characterization of the occurring groups can be found in table~\ref{list}.}
\label{results-groupsearch1}
\end{center}
\end{small}
\end{table}

\begin{table}
\vspace*{-10mm}
\begin{small}
\begin{center}
\renewcommand{\arraystretch}{1.2}
\begin{tabular}{|l|l|l|l|}
\hline
$G$ &
$\left( \left| v_1 \right|^2,\, \left| v_2 \right|^2,\,
\left|v_3 \right|^2 \right)$ &
$G_\ell$ &
$\left\langle \left\langle G_\ell,\, G_\nu \right\rangle \right\rangle$
\\ \hline
$[48, 30];\ [192, 182];\ [432, 260]$ &
$\left( 0,\, 1/2,\, 1/2\right)$ &
$\Z_4$ &
$[16, 3]$
\\ \hline
$[216, 95];\ [648, 259];\ [648, 260];\ [648, 266];\ [648, 563]$ &
$\left( 0,\, 1/2,\, 1/2\right)$ &
$\Z_6$ &
$[24, 10]$
\\ \hline
$[96, 64];\ [384, 568]$ &
$\left( 0,\, 1/2,\, 1/2\right)$ &
$\Z_4$ &
$[32, 11]$
\\ \hline
$[96, 65];\ [384, 571]$ &
$\left( 0,\, 1/2,\, 1/2\right)$ &
$\Z_8$ &
$[32, 5]$
\\ \hline
$[648, 266]$ &
$\left( 0,\, 1/2,\, 1/2\right)$ &
$\Z_3$ &
$[36, 12]$
\\ \hline
$[432, 260]$ &
$\left( 0,\, 1/2,\, 1/2\right)$ &
$\Z_{12}$ &
$[48, 21]$
\\ \hline
$[192, 186]$ &
$\left( 0,\, 1/2,\, 1/2\right)$ &
$\Z_{16}$ &
$[64, 29]$
\\ \hline
$[648, 266]$ &
$\left( 0,\, 1/2,\, 1/2\right)$ &
$\Z_6$ &
$[72, 30]$
\\ \hline
$[648, 563]$ &
$\left( 0,\, 1/2,\, 1/2\right)$ &
$\Z_{18}$ &
$[72, 10]$
\\ \hline
$[600, 179]$ &
$\left( 0,\, 1/2,\, 1/2\right)$ &
$\Z_5$ &
$[100, 14]$
\\ \hline
$[648, 259];\ [648, 260]$ &
$\left( 0,\, 1/2,\, 1/2\right)$ &
$\Z_9$ &
$[108, 24]$
\\ \hline
$[384, 568]$ &
$\left( 0,\, 1/2,\, 1/2\right)$ &
$\Z_8$ &
$[128, 67]$
\\ \hline
$[384, 581]$ &
$\left( 0,\, 1/2,\, 1/2\right)$ &
$\Z_{32}$ &
$[128, 131]$
\\ \hline
$[600, 179]$ &
$\left( 0,\, 1/2,\, 1/2\right)$ &
$\Z_{10}$ &
$[200, 31]$
\\ \hline
$[648, 259];\ [648, 260]$ &
$\left( 0,\, 1/2,\, 1/2\right)$ &
$\Z_{18}$ &
$[216, 58]$
\\ \hline\hline
$[216, 95];\ [648, 259];\ [648, 260];\ [648, 266];\ [648, 563]$ &
$\left( 1/4,\, 1/4,\, 1/2\right)$ &
$\Z_6$ &
$[72, 42]$
\\ \hline
$[648, 563]$ &
$\left( 1/4,\, 1/4,\, 1/2\right)$ &
$\Z_{18}$ &
$[216, 89]$
\\ \hline\hline
$[216, 95];\ [648, 259];\ [648, 260];\ [648, 266];\ [648, 563]$ &
$\left( 0,\, 1/4,\, 3/4\right)$ &
$\Z_6$ &
$[36, 12]$
\\ \hline
$[648, 563]$ &
$\left( 0,\, 1/4,\, 3/4\right)$ &
$\Z_{18}$ &
$[108, 24]$
\\ \hline
\end{tabular} 
\caption{In the first column,
the groups resulting from the search described in section~2.2
and of order smaller than 800.
In the second column,
the corresponding values of the $\left| v_k \right|^2$
($k = 1, 2, 3$).
The group $G_\ell$ is shown in the third column
and the smallest finite group having $G_\ell$ and $G_\nu$ as subgroups
is listed in the fourth column.
A characterization of the occurring groups can be found in table~\ref{list}.}

\label{results-groupsearchZn}
\label{results-groupsearch2}
\end{center}
\end{small}
\end{table}

\begin{table}
\begin{small}
\begin{center}
\renewcommand{\arraystretch}{1.2}
\begin{tabular}{|l|l|}
\hline
$G$ & $\left\langle \left\langle G_\ell,\, G_\nu \right\rangle \right\rangle$
\\ \hline
$[24, 12] \cong S_4 \cong \Delta(6\times 2^2)$ &
$[8, 3] \cong D_4$
\\
$[48,30] \cong A_4 \rtimes \Z_4$ &
$[12, 4] \cong D_6$
\\
$[60, 5] \cong A_5$ &
$[16, 3] \cong (\Z_4 \times \Z_2) \rtimes \Z_2$
\\
$[96, 64] \cong \Delta(6\times 4^2)$ &
$[16, 7] \cong D_8$
\\
$[96, 65] \cong A_4\rtimes\Z_8$ &
$[20, 4] \cong D_{10}$
\\
$[168, 42] \cong \Sigma(168) \cong \mathrm{PSL}(2,7)$ &
$[24, 6] \cong D_{12}$
\\
$[192, 182] \cong ((\Z_4 \times \Z_4)\rtimes\Z_3)\rtimes\Z_4$ &
$[24, 10] \cong \Z_3 \times D_4$
\\
$[192, 186] \cong A_4 \rtimes \Z_{16}$ &
$[24, 12] \cong S_4$
\\
$[216, 95] \cong \Delta(6\times 6^2)$ &
$[28,3] \cong D_{14}$
\\
$[384, 568] \cong \Delta(6\times 8^2)$ &
$[32, 5] \cong (\Z_8 \times \Z_2) \rtimes \Z_2$
\\
$[384, 571] \cong ((\Z_4 \times \Z_4)\rtimes \Z_3)\rtimes \Z_8$ &
$[32, 11] \cong (\Z_4 \times \Z_4) \rtimes \Z_2$
\\
$[384,581] \cong A_4 \rtimes \Z_{32}$ &
$[36, 12] \cong \Z_6 \times S_3$
\\
$[432,260] \cong ((\Z_6 \times \Z_6)\rtimes \Z_3)\rtimes \Z_4$ &
$[48, 21] \cong \Z_3 \times ((\Z_{4} \times \Z_2) \rtimes \Z_2)$
\\
$[600, 179] \cong \Delta(6\times 10^2)$ &
$[60, 5] \cong A_5$
\\
$[648, 259] \cong D_{18,6}^{(1)} \cong (\Z_{18}\times\Z_6)\rtimes S_3$ &
$[64, 29] \cong (\Z_{16} \times \Z_2) \rtimes \Z_2$
\\
$[648, 260] \cong ((\Z_{18}\times\Z_6)\rtimes \Z_3)\rtimes \Z_2$ &
$[72, 10] \cong \Z_9 \times D_4$
\\
$[648, 266] \cong ((\Z_6\times\Z_6\times\Z_3)\rtimes Z_3)\rtimes \Z_2$ &
$[72, 30] \cong \Z_3 \times ((\Z_6 \times \Z_2) \rtimes \Z_2)$
\\
$[648, 563] \cong ((\Z_{18}\times\Z_6)\rtimes \Z_3)\rtimes \Z_2$ &
$[72, 42] \cong \Z_3 \times S_4$
\\
$[864, 701] \cong \Delta(6\times 12^2)$ &
$[100, 14] \cong \Z_{10} \times D_5$
\\
$[1080, 260] \cong \Sigma(360\times 3)$ &
$[108, 24] \cong \Z_{18} \times S_3$
\\
$[1176, 243] \cong \Delta(6\times 14^2)$ &
$[128, 67] \cong (\Z_8 \times \Z_8) \rtimes \Z_2$
\\
 & $[128, 131] \cong (\Z_{32} \times \Z_2) \rtimes \Z_2$
\\
 & $[200, 31] \cong \Z_{5} \times ((\Z_{10} \times \Z_2) \rtimes \Z_2)$
\\
 & $[216, 58] \cong \Z_{9} \times ((\Z_6 \times \Z_2) \rtimes \Z_2)$
\\
 & $[216, 89] \cong \Z_9 \times S_4$
\\ \hline
\end{tabular} 
\caption{List of the groups appearing
in tables~\ref{results-groupsearch1} and~\ref{results-groupsearch2}.
Details on those groups in the left column
which are of order smaller than 512 can be found
in ref.~\cite{U3-512}.
The symbol $D_{18,6}^{(1)}$
denotes an $SU(3)$ subgroup of type D, \textit{cf.}\ ref.~\cite{Dgroups}.}
\label{list}
\end{center}
\end{small}
\end{table}


\newpage

\begin{thebibliography}{99}

\bibitem{lam0}
C.S.~Lam,
\textit{Symmetry of lepton mixing},
Phys.\ Lett.\ B {\bf 656} (2007) 193
[{\tt arXiv:0708.3665 [hep-ph]}];\\
%
C.S.~Lam,
\textit{Determining horizontal symmetry from neutrino mixing},
Phys.\ Rev.\ Lett.\ {\bf 101} (2008) 121602
[{\tt arXiv:0804.2622 [hep-ph]}];\\
%
C.S.~Lam,
\textit{The unique horizontal symmetry of leptons},
Phys.\ Rev.\ D {\bf 78} (2008) 073015
[{\tt arXiv:0809.1185 [hep-ph]}];\\
%
C.S.~Lam,
\textit{A bottom-up analysis of horizontal symmetry},
{\tt arXiv:0907.2206 [hep-ph]};\\
%
C.S.~Lam,
\textit{Leptonic mixing and group structure constants},
Phys.\ Rev.\ D {\bf 87} (2013) 053018
[{\tt arXiv:1301.3121 [hep-ph]}].

\bibitem{lam}
C.S.~Lam,
\textit{Finite symmetry of leptonic mass matrices},
Phys.\ Rev.\ D {\bf 87} (2013) 013001
[{\tt arXiv:1208.5527 [hep-ph]}].

\bibitem{lam1}
C.S.~Lam,
\textit{Horizontal symmetries $\Delta(150)$ and $\Delta(600)$},
Phys.\ Rev.\ D {\bf 87} (2013) 053012
[{\tt arXiv:1301.1736 [hep-ph]}].

\bibitem{ge}
S.-F.~Ge, D.A.~Dicus, and W.W.~Repko,
\textit{$Z_2$ symmetry prediction for the leptonic Dirac $CP$ phase},
Phys.\ Lett.\ B {\bf 702} (2011) 220
[{\tt arXiv:1104.0602 [hep-ph]}];\\
%
S.-F.~Ge, D.A.~Dicus, and W.W.~Repko,
\textit{Residual symmetries for neutrino mixing with a large
$\theta_{13}$ and nearly maximal $\delta_D$}, 
Phys.\ Rev.\ Lett.\  {\bf 108} (2012) 041801
[{\tt arXiv:1108.0964 [hep-ph]}].

\bibitem{yin}
H.-J.\ He and F.-R.\ Yin,
\textit{Common origin of $\mu$--$\tau$ and $CP$ breaking in neutrino
seesaw, baryon asymmetry, and hidden flavor symmetry}, 
Phys.\ Rev.\ D {\bf 84} (2011) 033009
[{\tt arXiv:1104.2654 [hep-ph]}].

\bibitem{hagedorn0}
R.~de~Adelhart~Toorop, F.~Feruglio, and C.~Hagedorn,
\textit{Discrete flavour symmetries in light of T2K},
Phys.\ Lett.\ B {\bf 703} (2011) 447
[{\tt arXiv:1107.3486 [hep-ph]}].

\bibitem{hagedorn2}
R.~de~Adelhart~Toorop, F.~Feruglio, and C.~Hagedorn,
\textit{Finite modular groups and lepton mixing},
Nucl.\ Phys.\ B {\bf 858} (2012) 437
[{\tt arXiv:1112.1340 [hep-ph]}].

\bibitem{he}
H.-J.\ He and X.-J.\ Xu,
\textit{Octahedral symmetry with geometrical breaking:
New prediction for neutrino mixing angle $\theta_{13}$
and $CP$ violation},
Phys.\ Rev.\ D {\bf 86} (2012) 111301
[{\tt arXiv:1203.2908 [hep-ph]}].

\bibitem{smirnov}
D.~Hernandez and A.Yu.~Smirnov, 
\textit{Lepton mixing and discrete symmetries}, 
Phys.\ Rev.\ D {\bf 86} (2012) 053014 
[{\tt arXiv:1204.0445 [hep-ph]}];\\
%
D.~Hernandez and A.Yu.~Smirnov, 
\textit{Discrete symmetries and model-independent patterns of lepton mixing}, 
Phys.\ Rev.\ D {\bf 87} (2013) 053005
[{\tt arXiv:1212.2149 [hep-ph]}];\\
%
D.~Hernandez and A.Yu.~Smirnov,
\textit{Relating neutrino masses and mixings by discrete symmetries},
Phys.\ Rev.\ D {\bf 88} (2013) 093007
[{\tt arXiv:1304.7738 [hep-ph]}].

\bibitem{feruglio}
F.~Feruglio, C.~Hagedorn, and R.~Ziegler,
\textit{Lepton mixing parameters from discrete and $CP$ symmetries},
J.\ High Energy Phys.\ {\bf 1307} (2013) 027
[{\tt arXiv:1211.5560 [hep-ph]}].

\bibitem{lindner}
M.~Holthausen, K.S.~Lim, and M.~Lindner,
\textit{Lepton mixing patterns from a scan of finite discrete groups},
Phys.\ Lett.\ B {\bf 721} (2013) 61
[{\tt arXiv:1212.2411 [hep-ph]}].

\bibitem{hu}
B.~Hu,
\textit{Neutrino mixing and discrete symmetries},
Phys.\ Rev.\ D {\bf 87} (2013) 033002
[{\tt arXiv:1212.2819 [hep-ph]}].

\bibitem{Walter}
W.~Grimus,
\textit{Discrete symmetries, roots of unity, and lepton mixing},
J.\ Phys.\ G {\bf 40} (2013) 075008
[{\tt arXiv:1301.0495 [hep-ph]}].

\bibitem{review-King}
S.F.~King and C.~Luhn,
\textit{Neutrino mass and mixing with discrete symmetry},
Rep.\ Prog.\ Phys.\  {\bf 76} (2013) 056201
[{\tt arXiv:1301.1340 [hep-ph]}].

\bibitem{joshipura}
A.S.~Joshipura and K.M.~Patel,
\textit{Horizontal symmetries of leptons with a massless neutrino},
Phys.\  Lett.\ B {\bf 727} (2013) 132
[{\tt arXiv:1306.1890 [hep-ph]}].

\bibitem{hagedorn1}
C.~Hagedorn, A.~Meroni, and L.~Vitale,
\textit{Mixing patterns from the groups $\Sigma \left( n \varphi \right)$},
J.\ Phys.\ A {\bf 47} (2014) 055201
[{\tt arXiv:1307.5308 [hep-ph]}].

\bibitem{Luhn}
P.~Ballett, S.F.~King, C.~Luhn, S.~Pascoli, and M.A.~Schmidt,
\textit{Testing atmospheric mixing sum rules at precision neutrino facilities},
{\tt arXiv:1308.4314 [hep-ph]}.

\bibitem{Meloni}
D.~Meloni,
\textit{Checking flavour models at neutrino facilities},
{\tt arXiv:1308.4578 [hep-ph]}.

\bibitem{Ballett}
P.A.~Ballett,
\textit{Probing leptonic flavour
with future long-baseline neutrino oscillation experiments},
Ph.D.~thesis, Durham University (2013).

\bibitem{Hanlon}
A.D.~Hanlon, S.-F.~Ge, and W.W.~Repko,
\textit{Phenomenological study of residual $\Z_2^s$
and $\overline{\Z}_2^s$ symmetries},
{\tt arXiv:1308.6522 [hep-ph]}.

\bibitem{Araki}
T.~Araki, H.~Ishida, H.~Ishimori, T.~Kobayashi, and A.~Ogasahara,
\textit{CKM matrix and flavor symmetries},
Phys.\ Rev.\ D {\bf 88} (2013) 096002
{\tt [arXiv:1309.4217 [hep-ph]]}.

\bibitem{branco}
G.C.~Branco and L.~Lavoura,
\textit{Rephasing-invariant parametrization of the quark mixing matrix},
Phys.\ Lett.\ B {\bf 208} (1988) 123.

\bibitem{GAP}
\textit{GAP --- Groups, Algorithms, Programming --- a System
for Computational Discrete Algebra}, in {\tt http://www.gap-system.org/}.

\bibitem{SmallGroups}
H.U.~Besche, B.~Eick, and E.~O'Brien,
\textit{The SmallGroups Library},
\\ in {\tt http://www.gap-system.org/Packages/sgl.html}.

\bibitem{sonata}
E.~Aichinger, F.~Binder, J.~Ecker, P.~Mayr, and C.~N\"obauer,
\textit{GAP package SONATA --- System of nearrings and their applications},
\\ in {\tt http://www.gap-system.org/Packages/sonata.html}
\\ and {\tt http://www.algebra.uni-linz.ac.at/Sonata/}.

\bibitem{hagedorn}
C.~Hagedorn and D.~Meloni,
\textit{$D_{14}$ --- A common origin of the Cabibbo angle
and the lepton mixing angle $\theta^l_{13}$},
Nucl.\ Phys.\ B {\bf 862} (2012) 691
[{\tt arXiv:1204.0715 [hep-ph]}].

\bibitem{varzielas}
I.~de~Medeiros~Varzielas and L.~Lavoura,
\textit{Golden ratio lepton mixing and nonzero reactor angle with $A_5$},
{\tt arXiv:1312.0215 [hep-ph]}.

\bibitem{fogli}
G.L.~Fogli, E.~Lisi, A.~Marrone, D.~Montanino, A.~Palazzo, and A.M.~Rotunno,
\textit{Global analysis of neutrino masses, mixings, and phases:
Entering the era of leptonic $CP$ violation searches},
Phys.\ Rev.\ D {\bf 86} (2012) 013012
[{\tt arXiv:1205.5254 [hep-ph]}].

\bibitem{U3-512}
P.O.~Ludl,
\textit{On the finite subgroups of U(3) of order smaller than 512},
J.\ Phys.\ A {\bf 43} (2010) 395204
[Erratum: \textit{ibid.}\ A {\bf 44} (2011) 139501]
[{\tt arXiv:1006.1479 [math-ph]}].

\bibitem{Dgroups}
P.O.~Ludl,
\textit{Comments on the classification of the finite subgroups of SU(3)},
J.\ Phys.\ A {\bf 44} (2011) 255204
[Erratum: \textit{ibid.}\ A {\bf 45} (2012) 069502]
[{\tt arXiv:1101.2308 [math-ph]}];\\
%
W.~Grimus and P.O.~Ludl,
\textit{On the characterization of the SU(3)-subgroups of type C and D},
{\tt arXiv:1310.3746 [math-ph]}.

\end{thebibliography}
\end{document}